\documentclass[aps,prl,showpacs,amsmath,amssymb,twocolumn]{revtex4}

\usepackage{pslatex}
\usepackage{graphicx}
\usepackage{dcolumn}
\usepackage{bm}
\usepackage{natbib}
\begin{document}

\title{Nearly-logarithmic decay in the colloidal hard-sphere system}
\author{M. Sperl}
\affiliation{Fachbereich Physik, Universit\"at Konstanz,
78457 Konstanz, Germany}
\date{\today}

\begin{abstract}

Nearly-logarithmic decay is identified in the data for the mean-squared
displacement of the colloidal hard-sphere system at the liquid-glass
transition [v. Megen \textit{et. al}, Phys.~Rev.~E \textbf{58}, 6073
(1998)]. The solutions of mode-coupling theory for the microscopic
equations of motion fit the experimental data well. Based on these
equations, the nearly-logarithmic decay is explained as the equivalent of
a $\beta$-peak phenomenon, a manifestation of the critical relaxation when
the coupling between of the probe variable and the density fluctuations is
strong.  In an asymptotic expansion, a Cole-Cole formula including
corrections is derived from the microscopic equations of motion, which
describes the experimental data for three decades in time.

\end{abstract}

\maketitle

The transition of a fluid to a glass is accompanied by a dramatic slowing
down of the dynamics as the system approaches the transition point
\cite{Kob2003}.  The correlation function for some variable $A$,
$\phi_A(t) = \langle A^*(t)A\rangle / \langle|A|^2\rangle$, with $<>$
denoting canonical averaging, decays rapidly typically within a time
increase by a factor of 10 in the normal fluid regime, while it stretches
over orders of magnitude in time close to a glass transition.
Simultaneously nontrivial dynamical features like power law decay and
unconventional scaling are observed. The characterization and explanation
of these signatures of the glass transition present challenges to
experiment, computer simulation and theory alike. One such feature is the
nearly-logarithmic decay of the correlation function which was discovered
recently in the orientational correlation function by optical-Kerr-effect
(OKE) measurements in molecular liquids \cite{Hinze2000}. The response
function measured in these experiments is proportional to the negative
time derivative of the correlation function, so a measured nearly-$t^{-1}$
decay is equivalent to a nearly-logarithmic relaxation in $\phi_A(t)$. In
a subsequent fit it was shown that the data can be welldescribed by a
schematic model within the mode-coupling theory for ideal glass
transitions (MCT) \cite{Goetze2004}. The nearly-logarithmic decay could be
interpreted as a $\beta$-peak phenomenon of the short-time critical
relaxation when rotation-translation coupling is sufficiently large.
Different from earlier suggestions, the presence of higher-order glass
transitions, like in systems with very short-ranged attractions
\cite{Sperl2003a}, could be ruled out as a cause for the
nearly-logarithmic decay in molecular systems.

Schematic models are truncated versions of the full equations of motion of
MCT that share the mathematical structure and universal properties of the 
latter but do not contain microscopic details \cite{Goetze1991b}. Schematic 
models involve fit parameters in the equations of motion for the correlation 
functions, what limits the interpretation of fits like in 
Ref.~\cite{Goetze2004} with respect to the microscopic origin of the observed 
features -- like the nearly-logarithmic decay -- that go beyond universal
predictions.  
In the full microscopic equations, these parameters are 
completely determined by the number density $\rho=N/V$ for a system of $N$ 
particles in a volume $V$, and the static structure factor $S_q$ which is given
by the interaction potential \cite{Bengtzelius1984,Goetze1991b,Hansen1986}. 
In the following, a system of hard spheres of diameter $d$ shall be discussed,
where the only external control parameter is the packing fraction $\varphi = 
\pi\rho d^3/6$. Predictions for the hard-sphere system (HSS) can be tested in 
colloidal suspensions, and the results for collective density fluctuations are 
reviewed in \cite{Megen1995}, demonstrating the universal laws of MCT and 
verifying certain characteristic parameters of the theory. The tagged particle 
dynamics has been analyzed recently in a computer-simulation study showing
good agreement with MCT \cite{Voigtmann2004}.

A particularly informative quantity to be studied in a system that shows
structural relaxation is the mean-squared displacement (MSD), which is
defined by $\delta r^2(t) = \langle|\vec{r}(t) - \vec{r}(0)|^2\rangle$
with $\vec{r}(t)$ denoting the position of a particle at time $t$. The MSD is 
studied frequently in computer simulation and was measured for nine decades in 
time in a colloidal suspension \cite{Megen1998}. 
In this paper, the MSD will be calculated for the HSS within MCT in order to
show, that for a state near the glass transition there is a window of two 
decades in time of nearly-logarithmic relaxation. This will be done by
identifying a $t^{-1}$ decay of the derivative of the MSD in complete analogy
to the procedure used for the analysis of the corresponding result for the
OKE data \cite{Hinze2000}. Second, by asymptotic solution of the MCT equations,
a modified Cole-Cole law will be derived for the description of the critical 
decay of the MSD. This analytical formula explains the nearly-logarithmic
regime and accounts for the critical decay for three
decades in time. Third, it will be shown that the MCT solutions at the
critical point describe the experimental result in \cite{Megen1998} for five 
orders of magnitude in time. Fourth, the full range of available experimental
data for the MSD is described reasonably by the numerical solutions of the MCT
equations going beyond the universal laws. Thus nearly-logarithmic decay laws
known from molecular systems as $\beta$-peak phenomenon are established in 
a colloidal system.

\begin{figure}[tb]
\includegraphics[width=\columnwidth]{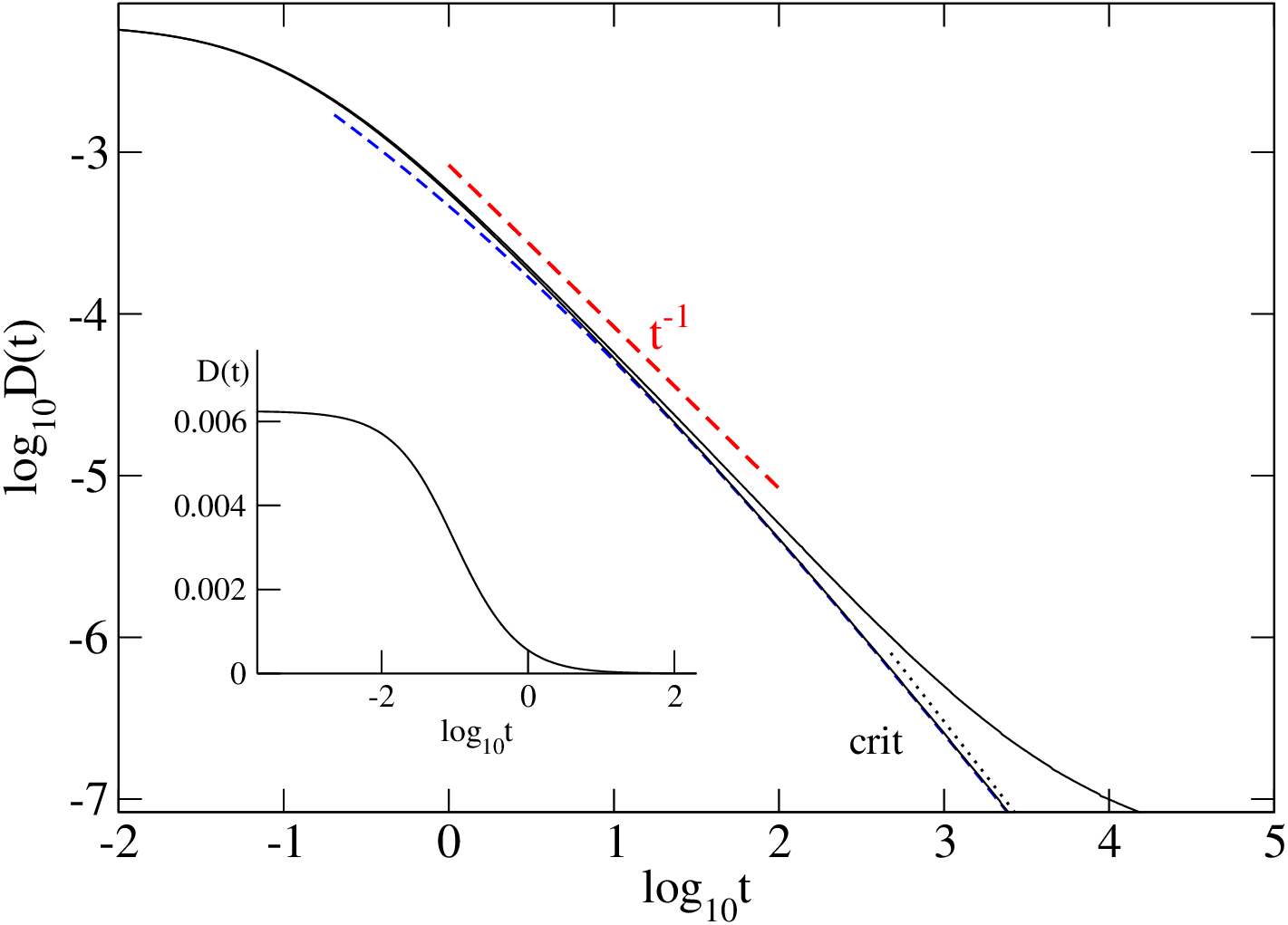}
\caption{\label{fig:MSDder}Time derivative of the mean-squared displacement 
$D(t)$ in 
the hard-sphere system for packing fractions $\varphi = \varphi^c = 0.5159$
and $\varphi = 0.5145$ (full lines from left to right). The upper dashed
line is proportional to $t^{-1}$. The lower dashed curve represents 
the approximation by Eq.~(\ref{eq:MSDML}). The dotted line displays the 
critical law $t^{-a-1}$. Here and in Fig.~\ref{fig:MSD} the unit of time is
fixed by the short-time diffusion coefficient $D^s_0/d^2=1/160$. The inset
shows $D(t)$ for $\varphi = \varphi^c$ on a linear scale for comparison.
}
\end{figure}

Slow dynamics within MCT originates from a singularity in the long-time 
behavior of the collective density fluctuations \cite{Bengtzelius1984}. Other
variables share that behavior if they are coupled to the density fluctuations.
While the coupling to rotational motion was modeled by an empirical parameter
in \cite{Goetze2004}, the MSD for a colloidal system is given be a microscopic
equation of motion by \cite{Fuchs1998},
\begin{subequations}\label{eq:MSD:MCT}
\begin{equation}\label{eq:MSD:MCT:B}
\delta r^2(t) +D^s_0\int_0^t\,dt'\, m^{(0)}(t-t')\delta r^2(t')= 6 D^s_0 
t\,,
\end{equation}
with the short-time diffusion coefficient $D^s_0$. The memory kernel 
$m^{(0)}(t)$ is determined by the static structure of the liquid and 
collective and single density correlation functions. The structure factor is
calculated using the Percus-Yevick approximation (PYA) \cite{Hansen1986}, 
details of the numerical solution are outlined elsewhere \cite{Fuchs1998}. 
For $\varphi$ below the critical point $\varphi^c = 0.5159$, the long-time 
solution of Eq.~(\ref{eq:MSD:MCT:B}) is diffusive, $\delta r^2(t)=6D^st$, 
with a long-time diffusion coefficient $D^s$ depending on the distance 
from the critical point. For $\varphi \geqslant \varphi^c$, the long-time 
solution is arrested at a plateau $\delta r^2(t)=6r_s^2$, with a localization 
length $r_s$. The derivative of the MSD can be interpreted as a time-dependent
diffusion coefficient, $6D(t)=d[\delta r^2(t)/d^2]/dt$, that goes to zero as
the glass transition is approached \cite{Segre1996}. $D(t)$ calculated from
Eq.~(\ref{eq:MSD:MCT:B}) is plotted for $\varphi = 0.5145$ and $\varphi = 
\varphi^c$ as full lines in Fig.~\ref{fig:MSDder}. The dynamics when 
approaching the critical point 
shares more and more of the relaxation at $\varphi^c$ as seen for $\varphi 
= 0.5145$ for $\log_{10}t<1$. This portion of the dynamics is therefore 
referred to as critical relaxation. For times exceeding the critical 
relaxation, the dynamics crosses over to the long-time diffusion as seen 
for $\varphi = 0.5145$ around $\log_{10}t \approx 2$.
The most significant finding in connection with Fig.~\ref{fig:MSDder} is
the appearance of a window in time after the transient where the dynamics
follows closely a $t^{-1}$ law for $0 \leqslant \log_{10}t \leqslant 2$. A 
comparison of the solution for $\varphi=0.5145$ with the critical decay 
shows that the major fraction of the $t^{-1}$ decay is part of the 
critical relaxation. This decay reflects the same behavior as found in 
molecular liquids, moreover, the decay observed in Fig.~\ref{fig:MSDder} 
is closer to $t^{-1}$ than for most of the OKE data where $t^{-x}$ is 
seen with $x$ ranging from 0.8 to 1.15 \cite{Hinze2000,Goetze2004}. The 
long-time decay at the critical point is a $t^{-a}$-law with a=0.312 for the 
HSS. This law is shown as dotted straight line of
slope $-(1+a)$. It accounts for the critical decay only for $\log_{10} t
\geqslant 3$, is preceeded by the $t^{-1}$ decay, and has no relevance for 
$\varphi=0.5145$.

\begin{figure}[tb]
\includegraphics[width=\columnwidth]{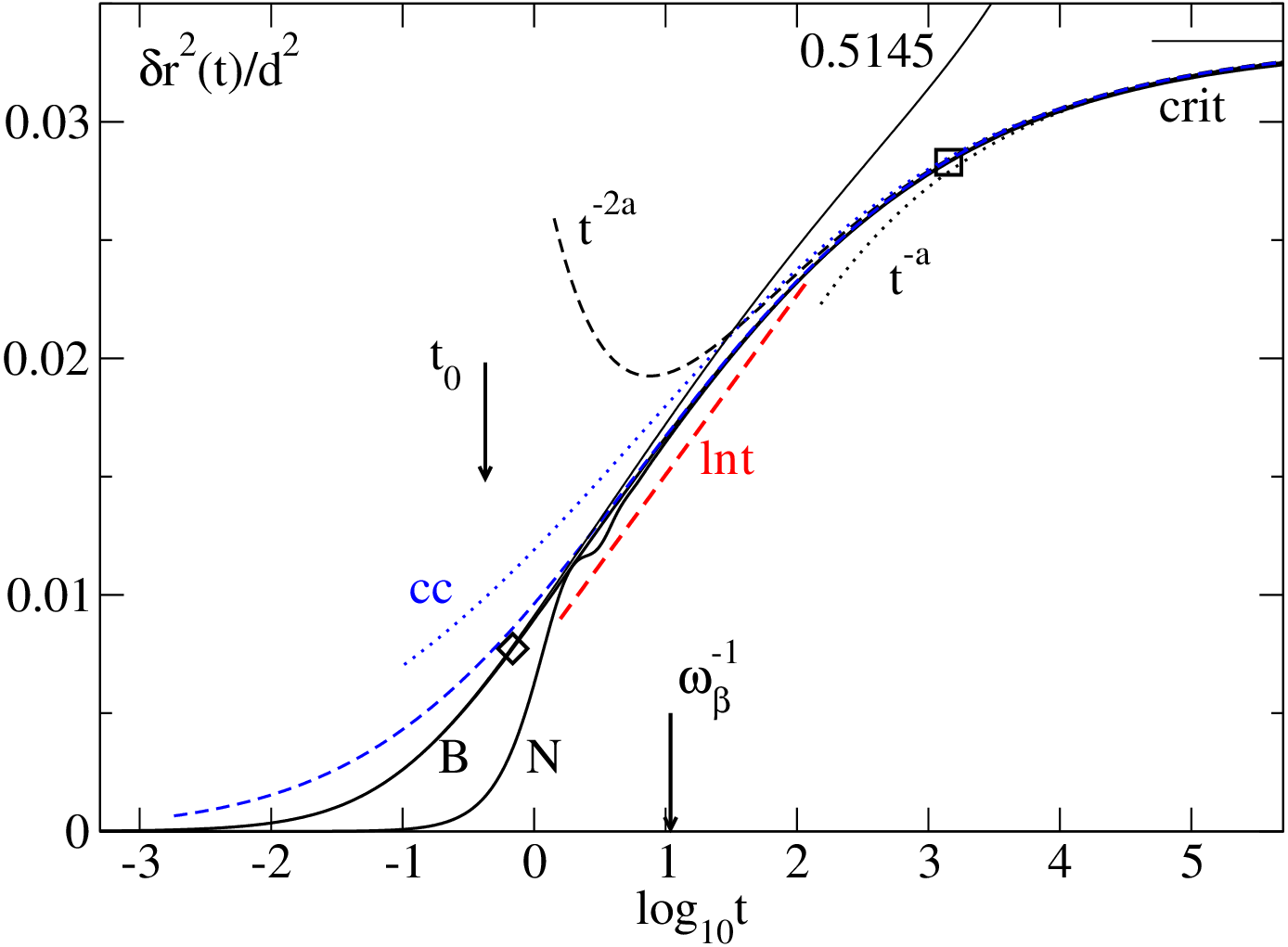}
\caption{\label{fig:MSD}Mean-squared displacement at the critical point in 
the hard-sphere system for Newtonian (N) and Brownian (B) dynamics (full 
curves). The short horizontal line displays the plateau $6 r_s^{c\,2}$.
Dashed lines show the asymptotic approximations by
Eqs.~(\ref{eq:MSD:crit_asy}) (labeled $t^{-2a}$) and (\ref{eq:MSDML}). The 
leading-order approximations are shown dotted for Eq.~(\ref{eq:MSD:crit_asy}) 
($t^{-a}$) and Eq.(\ref{eq:MSDML}) (cc). The nearly-logarithmic 
relaxation is shown by the straight dashed line labeled $\ln t$. The diamond
and the square indicate 10\% deviations of the critical solution B from the 
approximation by Eq.~(\ref{eq:MSDML}) and the solution for $\varphi = 
0.5145$, respectively.
}
\end{figure}

The nearly-logarithmic decay is found adjacent to the transient 
for the Brownian colloidal system as for the Newtonian molecular systems 
\cite{Goetze2004}. For the HSS also Newtonian dynamics can be considered and
the equivalent of Eq.~(\ref{eq:MSD:MCT:B}) reads 
\cite{Chong2001b},
\begin{equation}\label{eq:MSD:MCT:N}
\partial_t\delta r^2(t) + v_s^2\int_0^t\,dt'\, m^{(0)}(t-t')\delta 
r^2(t')= 6 v_s^2 t\,,
\end{equation}
with the thermal velocity $v_s^2$ for a tagged particle.
\end{subequations}
Figure~\ref{fig:MSD} shows the solution for the MSD for Newtonian and 
Brownian dynamics for $\varphi = \varphi^c$. Both
solutions can be matched at long times and they overlap down to $\delta r^2(t)
\approx 0.01$ defining the beginning of the regime for structural relaxation
at $\log_{10}t \approx 0$ \cite{Chong2001b}. The nearly-logarithmic decay is 
shown as nearly-linear increase of the MSD with $\log t$ outside the transient 
within the same time interval as marked by the dashed line in 
Fig~\ref{fig:MSDder}.

To further analyze the window of nearly-logarithmic relaxation and in an 
attempt to distinguish it from a further approximation to a power law with 
very small exponent like in \cite{Tokuyama2002}, asymptotic 
expansions shall be applied. At the critical point, the MSD can be 
expanded in power laws and reads \cite{Fuchs1998},
\begin{equation}\label{eq:MSD:crit_asy}
\delta r^2 (t) /6 = r_s^{c\;2} - h_{MSD}  (t/t_0)^{-a} + h_{MSD}
K_{MSD} (t/t_0)^{-2a}\,,
\end{equation}
where the first two terms on the RHS constitute the leading-order result.
The values for the HSS are $r_s^{c} = 0.0746$, $h_{MSD} = 0.0116$, 
$K_{MSD} = -1.23$, $t_0 = 0.425$, $\lambda = 0.735$.
Leading and next-to-leading order result of Eq.~(\ref{eq:MSD:crit_asy}) 
are shown dotted and dashed in Fig.~\ref{fig:MSD} and describe the full 
solutions only for $t\gtrsim 1000$ and $t\gtrsim 100$, respectively. As will
be shown below, the leading order fails to apply in the 
experimentally relevant window, while including the correction still does 
not explain more than one decade of structural relaxation.

In the following, a new asymptotic solution is derived which is 
based on the expansion of the memory kernel. For long times, the
equations of motion for the MSD are the same for both Newtonian and
Brownian dynamics \cite{Goetze1991b,Chong2001b}, and can be represented 
with the modified Laplace transform ${\cal S}[f(t)](z)=iz\int_0^\infty 
dt\,\exp[izt]f(t)$ as
\begin{equation}\label{eq:struct_rel:MSD}
{\cal S}[\delta r^2(t)](z) = \frac{6}{{\cal S}[m^{(0)}(t)](z)}\,.
\end{equation}
An asymptotic expansion for the memory kernel equivalent to the one in
Eq.~(\ref{eq:MSD:crit_asy}) is given by $ m^{(0)}(t) = f^c_{m^{(0)}} +
h_{m^{(0)}} (t/t_0)^{-a} + h_{m^{(0)}} K_{m^{(0)}} (t/t_0)^{-2a}$, with
$f^c_{m^{(0)}}=1/r_s^{c\,2}$, $h_{m^{(0)}}=h_{MSD}/r_s^{c\,4}$,
$K_{m^{(0)}}= K_{MSD}+\lambda h_{MSD}/r_s^{c\,2}$, and
$\lambda=\Gamma(1-a)^2/\Gamma(1-2a)$ with the Euler Gamma function
$\Gamma(x)$. In the Laplace domain, this can be written with a
characteristic frequency $\omega_\beta$ as ${\cal S}[m^{(0)}(t)](z) =
f^c_{m^{(0)}}\left\{ 1+\left(-iz/\omega_\beta\right)^a + K^{cc}_{MSD}
\left(-iz/\omega_\beta\right)^{2a} \right\}$, where
$K^{cc}_{MSD}=f^c_{m^{(0)}}K_{m^{(0)}}/(h_{m^{(0)}}\lambda)$. Hence,
Eq.~(\ref{eq:struct_rel:MSD}) reads up to next-to-leading order
\begin{equation}\label{eq:MSDML}
{\cal S}[\delta r^2(t)](z) = \frac{6r_s^{c\;2}}{1+(-iz/\omega_\beta)^a
+K^{cc}_{MSD}(-iz/\omega_\beta)^{2a}}\,,
\end{equation} 
and the characteristic frequency is given by
$\omega_\beta = \frac{1}{t_0}
\left[\frac{r_s^{c\,2}}{h_{MSD}}\frac{1}{\Gamma(1-a)}\right]^{1/a}$.
For the HSS we get $t_0\omega_\beta = 0.03895$. 
The leading-order result in Eq.~(\ref{eq:MSDML}) is obtained for $K^{cc}_{MSD}
= 0$, and is known as Cole-Cole law. For different variables a similar result
was obtained before \cite{Goetze1989c,Goetze2004}. While the correction in
Eq.~(\ref{eq:MSD:crit_asy}) is comparably large, $K_{MSD} = -1.23$, the
analogous result for the memory kernel is only a fraction of it,
$K_{m^{(0)}} = 0.30$. The correction in Eq.~(\ref{eq:MSDML}) is even
smaller, $K^{cc}_{MSD} = 0.196$. As the corrections determine the range of
validity for the leading terms of the asymptotic expansions, the
leading-order result of Eq.~(\ref{eq:MSDML}) is superior
to the leading-order result of Eq.~(\ref{eq:MSD:crit_asy}) as seen in 
Fig.~\ref{fig:MSD}, where already the leading order of
Eq.~(\ref{eq:MSDML}) describes qualitatively the complete relaxation.  
Including the correction explains the complete window of structural
relaxation as seen by the diamond marking a 10\% deviation of the
approximation by Eq.~(\ref{eq:MSDML}) from the solution of 
Eq.~(\ref{eq:MSD:MCT:B}). A comparison in Fig.~\ref{fig:MSDder} shows that
Eq.~(\ref{eq:MSDML}) also covers the complete regime of nearly-logarithmic 
decay. For $\varphi = 0.5145$, $D(t)$ in Fig.~\ref{fig:MSDder} and the MSD in 
Fig.~\ref{fig:MSD} is described by Eq.~(\ref{eq:MSDML}) for $\log_{10} t 
\lesssim 3$. The crossover from the critical decay to the long-time diffusion 
yields an extended window in time where $\ln t$ fits the solution.

\begin{figure}[tb]
\includegraphics[width=\columnwidth]{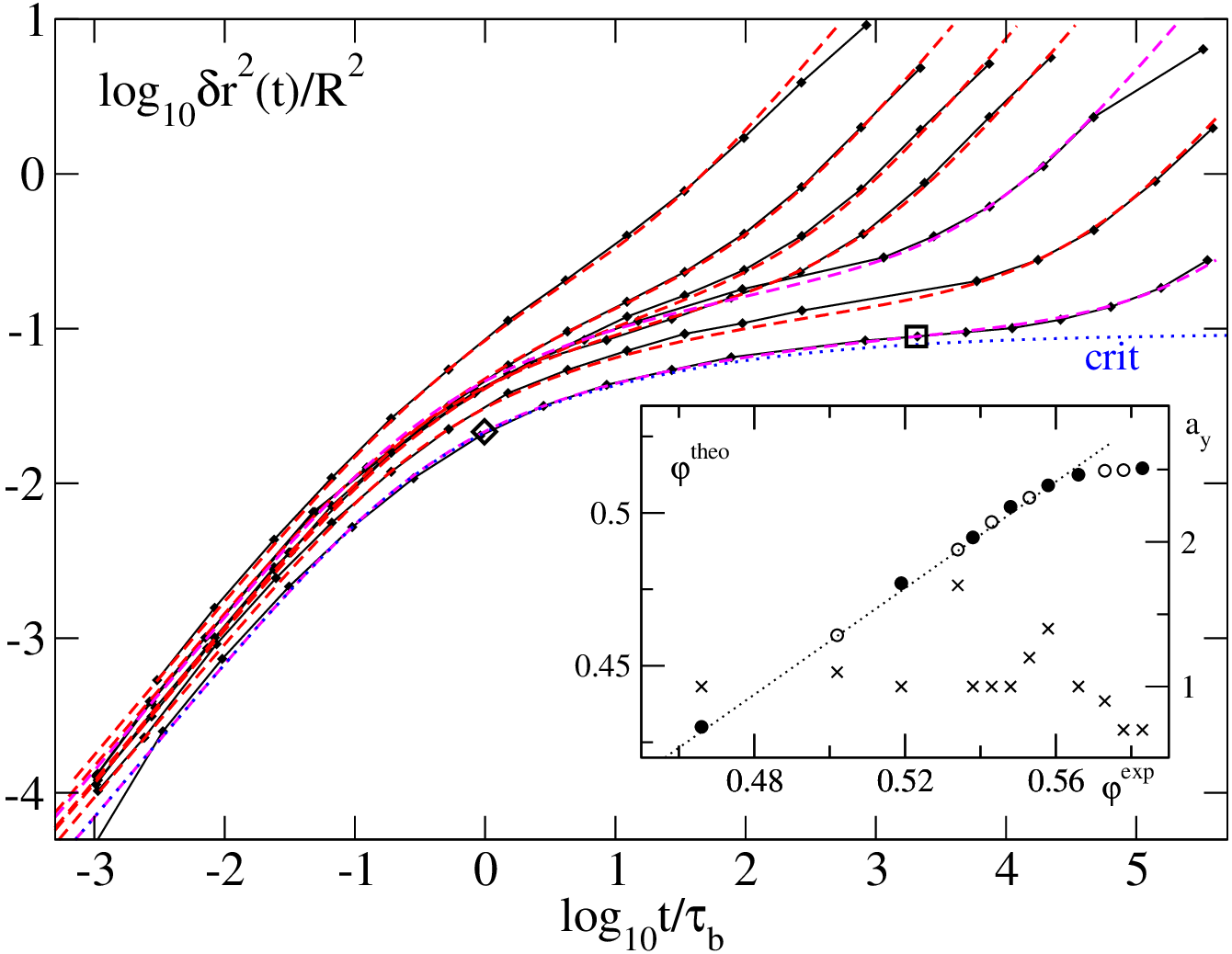}
\caption{\label{fig:MSDfit}Fit of the mean-squared-displacement data from
\cite{Megen1998} (diamonds and full curves) by the solutions of 
mode-coupling theory (dashed). The dotted curve displays the theoretical 
result at the critical point and is equivalent to curve B in 
Fig.~\ref{fig:MSD}. The square indicates a 10\% deviation between the critical
relaxation and the fit for $\varphi^\text{exp}=0.583$. The open diamond is the
same as in Fig.~\ref{fig:MSD}.
The inset shows the mapping of the experimental packing fractions
$\varphi^\text{exp}$ onto theoretical values $\varphi^\text{theo}$ on the left
axis as closed and open circles for curves shown or left out, respectively.
The dotted line shows the function $\varphi^\text{theo} = 0.87\,
\varphi^\text{exp} + 0.023$. The right axis displays the vertical scaling
factor by crosses.
}
\end{figure}

Figure~\ref{fig:MSDfit} shows the data from \cite{Megen1998} in units of the
particle radius $R=d/2$ and the Brownian time scale $\tau_b = R^2/(6D^s_0)$ 
together with the solutions of Eq.~(\ref{eq:MSD:MCT:B}) for 
different values of $\varphi$. The agreement is rather satisfactory
considering that all coupling parameters are fixed microscopically. Three
adjustments need to be made to match experimental and theoretical curves.
First, the short-time diffusion coefficient sets an
overall time scale that needs to be accounted for. The theoretical
solutions are matched to the data for $\log_{10}t\lesssim -1$. As the
experimental curves do not fall on top of each other in that regime, the
chosen time scale also needs to vary by a factor up to 2. 
Second, with $\varphi^c\approx 0.58$ \cite{Megen1998,Simeonova2004}, the
experimental glass transition in the HSS is found at approximately 12\%
higher packing fraction than predicted in
Refs.~\cite{Bengtzelius1984,Fuchs1998}. This discrepancy could be reduced
to below 10\% by using the actual structure factor rather than the PYA
in a computer simulation \cite{Foffi2004}, but a remaining error from the 
mode-coupling approximation has to be accepted. This is similar to other 
approaches fitting the data \cite{Tokuyama2003}. Therefore, theoretical 
values for $\varphi$ are mapped on experimental ones by comparing the 
diffusivity in experiment and theory. The result of this procedure is 
shown in the inset of Fig.~\ref{fig:MSDfit} by circles. The mapping is 
fitted by $\varphi^\text{theo} = 0.87\, \varphi^\text{exp} + 0.023$, which 
is remarkably similar to what was found from a computer simulation of the 
HSS \cite{Voigtmann2004}, and extrapolates to a critical density 
$\varphi^c\approx 0.57$. The highest values of $\varphi$ do not follow 
that linear mapping due to ageing for $\log_{10}t/\tau_b\gtrsim 4$ 
\cite{Simeonova2004,Megen1998}. The dynamics before ageing sets in, and 
hence the regime for nearly-logarithmic 
relaxation, is not affected significantly by ageing \cite{Simeonova2004}.
Third, as the actual transition takes place at a higher packing fraction
than theory predicts, a smaller localization length and hence a vertical shift 
of the curves by a prefactor $a_y$ -- shown by crosses in the inset of 
Fig.~\ref{fig:MSDfit} -- can be expected. Surprisingly, many curves 
are described well using $a_y=1$. The variation of $a_y$ yields an error 
estimate for the localization length of around 20\% for the
localization length as compared to the theoretical results. If only data
close to the glass transition is considered, the data show indeed a
smaller localization length than predicted. After these adjustments, 
Eq.~(\ref{eq:MSD:MCT:B}) fits the data for different packing fractions
over the entire experimental range of up to nine orders of magnitude in 
time.

The curve $\varphi^\text{exp} = 0.583$ is fitted by $\varphi^\text{theo} =
0.5146$, and in addition, the critical relaxation for $\varphi^\text{theo}
= \varphi^c$ is matched with the same prefactors and shown dotted. The data 
follow closely the critical relaxation for $t/\tau_b\lesssim 2000$. The two 
theoretical curves correspond to the ones shown in Figs.~\ref{fig:MSDder} 
and~\ref{fig:MSD}, thus establishing the existence of a window of 
nearly-logarithmic relaxation in the experimental data.

In conclusion, it is shown in Fig.~\ref{fig:MSDfit} that mode-coupling
theory (MCT) is able to describe the mean-squared displacement (MSD) in a
colloidal hard-sphere system (HSS) \cite{Megen1998} rather accurately. This
implies consistency with both universal laws of MCT and non-universal MCT
predictions for the HSS. Outside the transient, the data exhibit a window 
of two orders of magnitude in time where the relaxation resembles 
nearly-logarithmic decay or a $\beta$-peak phenomenon, cf. 
Fig.~\ref{fig:MSDder}. The latter is found independent of the underlying 
Brownian dynamics also for Newtonian dynamics, cf. Fig.~\ref{fig:MSD}, and 
is distinguished from an accidental crossover 
by an asymptotic expansion relating it to a $\beta$-peak, cf. 
Eq.~(\ref{eq:MSDML}). Such a dynamical feature was discovered in molecular 
systems \cite{Hinze2000} and interpreted as a consequence of strong
translation-rotation coupling \cite{Goetze2004}. In addition, similar
nearly-logarithmic decays were also observed in nematic liquid crystals 
\cite{Cang2003b}. For the hard-sphere system (HSS), while clearly lacking 
rotational degrees of freedom, the MSD constitutes another variable that is
coupled strongly to the density fluctuations if the tagged particle is
sufficiently large. It can be inferred from Ref.~\cite{Fuchs1998} that 
$m^{(0)}(t)$ in Eq.~(\ref{eq:MSD:MCT}) becomes smaller when the size of the
tagged particle is chosen smaller than the particles in the host fluid. A
larger particle in turn experiences a stronger coupling of its MSD to the
collective dynamics. Hence, the counterintuitive finding in \cite{Goetze2004},
that the dynamics of a coupled variable can deviate further from the
underlying collective dynamics as the coupling is increased, also applies to 
the MSD of the HSS for increasing the diameter of the particle for which the
MSD is measured. 
The empirical nearly-logarithmic law is part of the 
critical relaxation and does not change slope when observed
sufficiently close to the transition point. This is in clear contrast to the 
logarithmic laws in the vicinity of higher-order glass-transition
singularities \cite{Sperl2003a} and can be used to distinguish both relaxation
features.
Hydrodynamic interactions present in colloidal suspensions are not included in
Eq.~(\ref{eq:MSD:MCT}) but are known to change the short-time dynamics in MCT
\cite{Fuchs1999c}. The quality of the data fit and the validity of the
nearly-logarithmic law are therefore surprising and rule out a significant
contribution of hydrodynamic interactions on the dynamics of the MSD 
within MCT as shown in Fig.~\ref{fig:MSDfit}, which is in contrast to 
other approaches \cite{Tokuyama2002,Tokuyama2003}. This can be 
rationalized by the strong coupling of 
the MSD to the collective dynamics causing the MSD for a larger particle to be 
slower than for a smaller particle \cite{Fuchs1998}. For the slowest
collective relaxations only small deviations by hydrodynamic interactions have
been found \cite{Fuchs1999c}, a result that apparently seems to apply also to 
the slow coupled variable MSD. 

Equation~(\ref{eq:MSDML}) improves the understanding of glassy dynamics in
three aspects. First, it comprises a Cole-Cole that is derived rigorously
from a microscopic equations of motion and therefore provides a foundation
for data interpretations as in Ref.~\cite{Goetze2004}.  Second, it closes 
a gap in the understanding of the MSD close to the glass transition
\cite{Chong2001b}. Together with the results from Ref.~\cite{Fuchs1998},
MCT provides analytical descriptions of the dynamics in all subsequent
windows in time up to small crossover regions: The initial ballistic or
Brownian dynamics is followed by a critical relaxation described by
Eq.~(\ref{eq:MSDML}), before a plateau $6 r_s^{c\;2}$ is reached. After
that plateau, the crossover to the long-time diffusion starts with a
von~Schweidler law $t^{-b}$ where $b=0.583$ \cite{Fuchs1998}. Third, 
precise meaning is given to the finding that strongly coupled variables do 
not necessarily show the same dynamics.

This work was stimulated by discussions with H.Z.~Cummins, M.~Fuchs and
W.~G\"otze, and supported by the DFG SFB 513 and the DFG Grant No. SP~714/3-1.

\vspace{-5mm}

\end{document}